\documentclass[prl,twocolumn,superscriptaddress]{revtex4}
\usepackage{amsmath, amssymb,graphicx,color,bm,epstopdf}
\usepackage[dvipsnames]{xcolor}
\usepackage{comment}

\begin{document}

\title{Circuit Implementation of a Four-Dimensional Topological Insulator}

\author{You Wang}
\affiliation{Division of Physics and Applied Physics, School of Physical and Mathematical Sciences, Nanyang Technological University, Singapore 637371, Singapore}

\author{Hannah M. Price}
\affiliation{School of Physics and Astronomy, University of Birmingham, Edgbaston, Birmingham B15 2TT, United Kingdom}

\author{Baile Zhang}
\affiliation{Division of Physics and Applied Physics, School of Physical and Mathematical Sciences, Nanyang Technological University, Singapore 637371, Singapore}
\affiliation{Centre for Disruptive Photonic Technologies, Nanyang Technological University, Singapore 637371, Singapore}

\author{Y.~D.~Chong}
\affiliation{Division of Physics and Applied Physics, School of Physical and Mathematical Sciences, Nanyang Technological University, Singapore 637371, Singapore}
\affiliation{Centre for Disruptive Photonic Technologies, Nanyang Technological University, Singapore 637371, Singapore}

\begin{abstract}
The classification of topological insulators predicts the existence of high-dimensional topological phases that cannot occur in real materials, as these are limited to three or fewer spatial dimensions.  We use electric circuits to experimentally implement a four-dimensional (4D) topological lattice.  The lattice dimensionality is established by circuit connections, and not by mapping to a lower-dimensional system.  On the lattice's three-dimensional surface, we observe topological surface states that are associated with a nonzero second Chern number but vanishing first Chern numbers.  The 4D lattice belongs to symmetry class AI, which refers to time-reversal-invariant and spinless systems with no special spatial symmetry.  Class AI is topologically trivial in one to three spatial dimensions, so 4D is the lowest possible dimension for achieving a topological insulator in this class.  This work paves the way to the use of electric circuits for exploring high-dimensional topological models.
\end{abstract}

\maketitle

\textit{Introduction.---}Topological insulators are materials that are insulating in the bulk but host surface states protected by nontrivial topological features of their bulk bandstructures \cite{Hasan_2010, Qi_2011}.  They are classified according to symmetry and dimensionality~\cite{Altland1997, Kitaev_2009, Schnyder_2008, Ryu_2010, Chiu_2016}, with each class having distinct and interesting properties.  The celebrated two-dimensional Quantum Hall (2DQH) phase \cite{Klitzing_1980}, for instance, has topological edge states that travel unidirectionally on the one-dimensional (1D) edge, whereas three-dimensional (3D) topological insulators based on spin-orbit coupling have surface states that act like massless 2D Dirac particles.  The classification of topological insulators contains hypothetical high-dimensional phases \cite{Altland1997} that cannot be realized with real materials, since electrons only move in one, two, or three spatial dimensions.  These include several types of four-dimensional Quantum Hall (4DQH) phases, which are characterised by a 4D topological invariant called the second Chern number and exhibit a much richer phenomenology than the 2DQH phase \cite{avron1988topological, frohlich2000new, Zhang_2001, sugawa2018second}.  In recent years, topological phases have been implemented in a range of engineered systems including cold atom lattices~\cite{Cooper_2019}, photonic structures \cite{Ozawa_2019}, acoustic and mechanical resonators~\cite{Yang_2015, Huber2016}, and electric circuits \cite{Ningyuan_2015, Albert_2015, Hadad2018, Lee_2018, Imhof_2018, Luo_2018, Ezawa_2018, Wang2019, Lu_2019, Serra_Garcia_2019, Helbig_2019, Hofmann_2019}.  Some of these platforms can realise lattices that are hard to achieve in real materials, raising the intriguing prospect of using them to create high-dimensional topological insulators.  Although there have been demonstrations of ``topological pumps'' that map 4D topological lattice states onto lower-dimensional systems \cite{thouless1983quantization, Kraus_2013, Petrides_2018, lee2018electromagnetic}, there has been no experimental realisation of a 4D topological insulator with protected surface states on a 3D surface.

Here, we describe the implementation of a 4DQH phase using electric circuits to access higher dimensions.  Since electric circuits are defined in terms of lumped (discrete) elements and their interconnections, lattices with genuine high-dimensional structure can be explicitly constructed by applying the appropriate connections~\cite{Juki_2013, ezawa2019electric, li2019emergence}.  In this way, we experimentally implement a 4D lattice hosting the first realisation of a Class AI topological insulator \cite{Schnyder_2008, Ryu_2010}, which has no counterpart in three or fewer spatial dimensions.

In the symmetry-based classification of topological phases~\cite{Altland1997, Kitaev_2009, Chiu_2016, Ryu_2010, Schnyder_2008}, Class AI includes time-reversal (T) symmetric, spinless systems that are not protected by any special spatial symmetries.  Whereas the 2DQH phase is tied to nontrivial values of the first Chern number, which requires T-breaking~\cite{Thouless_1982}, 4DQH phases rely on the second Chern number, which does not~\cite{avron1988topological, frohlich2000new, Zhang_2001, sugawa2018second}.  Even though the Class AI conditions are ubiquitous~\cite{Cooper_2019, Ozawa_2019}, the class is topologically trivial in one to three dimensions~\cite{Altland1997, Kitaev_2009, Chiu_2016, Ryu_2010, Schnyder_2008}.  Hence, realising a Class AI topological insulator requires going to at least 4D.  We focus on a theoretical 4D lattice model recently developed by one of the authors \cite{Price2018}, which exhibits a nonzero second Chern number with vanishing first Chern numbers.  Hence, we obtain the first observations of topological surface states that are intrinsically tied to 4D band topology, with no connection to lower-dimensional topological invariants.

The present approach, based on circuit connections, is distinct from other recently-investigated methods for accessing higher-dimensional models.  One of the alternatives involves manipulating internal degrees of freedom, such as oscillator modes, to act as synthetic dimensions \cite{Boada_2012, Celi_2014, Price_2015, Mancini_2015, Stuhl_2015, Ozawa_2016, Price_2016, Yuan_2016, Ozawa_2017, Price_2017, yuan2018synthetic, price2019synthetic, lustig2019photonic, ozawa2019topological, dutt2020single}.
Although there have been theoretical proposals for using synthetic dimensions to build 4D topological lattices~\cite{Price_2015, Ozawa_2016}, all experiments so far have been limited to 1D and 2D~\cite{ozawa2019topological}.  Another approach involves adiabatic topological pumping schemes, which map high-dimensional models onto lower-dimensional setups by replacing spatial degrees of freedom with tunable parameters~\cite{thouless1983quantization, Kraus_2013, Petrides_2018, lee2018electromagnetic}.  As mentioned above, 2D topological pumps based on cold atoms and photonics have recently been used to explore Class A (T-broken) 4DQH systems \cite{Kraus_2013, Lohse_2018, Zilberberg_2018}. However, topological pumps have the drawback of being inherently limited to probing specific quasi-static solutions of a high-dimensional system, without realising a genuinely high-dimensional lattice.  Moreover, in those experiments the second Chern number in 4D is not truly independent of the first Chern numbers in 2D, which are nonzero.


Our 4D lattice is implemented using electric circuits with carefully chosen capacitive and inductive connections.   The lattice model has two topologically distinct phases: a 4DQH phase and a conventional insulator, with the choice of phase governed by a parameter $m$ that maps to certain combinations of capacitances and inductances.  Using impedance measurements that are equivalent to finding the local density of states (LDOS), we show that the 4DQH phase hosts surface states on the 3D surface, while the conventional insulator phase has only bulk states.  By varying the driving frequency, we show that the topological surface states span a frequency range corresponding to a bulk bandgap, as predicted by theory.  Our experimental results also agree well with circuit simulations.  This work demonstrates that electric circuits are a flexible and practical way to realise higher-dimensional lattices, paving the way for the exploration of other previously-inaccessible topological phases.

\begin{figure}
\includegraphics[width=\columnwidth]{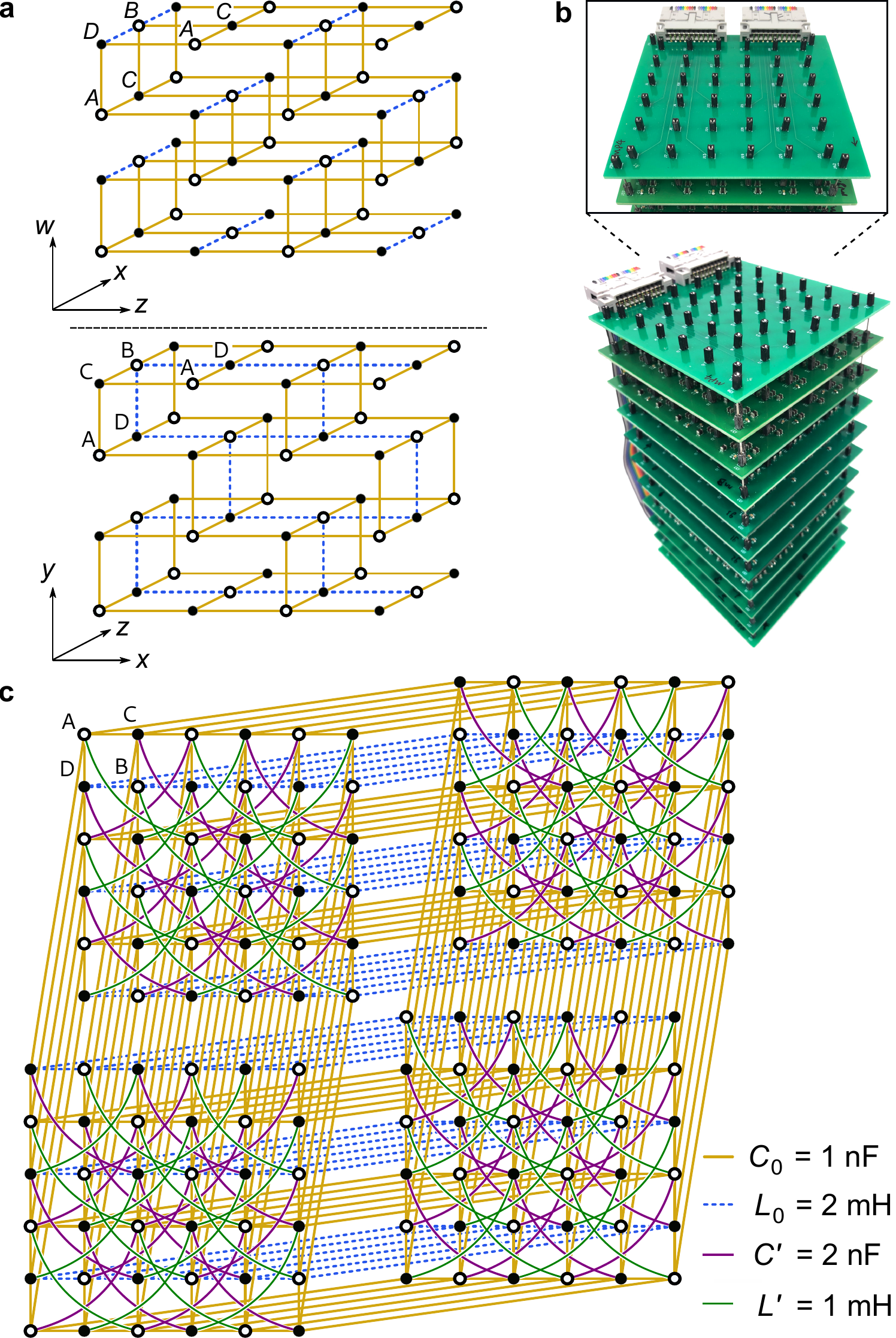}
\caption{Model of the 4D Quantum Hall lattice and its circuit implementation. (a) Schematic of the 4D tight-binding model. Each unit cell consists of four sites labelled A-D.  Hollow and filled circles respectively denote positive ($m$) and negative ($-m$) on-site masses, while yellow solid lines and blue dashes respectively denote positive ($J$) and negative ($-J$) hoppings. (c) Long-range hoppings of the tight-binding lattice. (b) Photographs of the circuit.  (c) Schematic of the circuit; positive (negative) masses are realised by capacitors (inductors) connecting the sites to ground, and hoppings are realised using capacitors or inductors connecting different sites.}
\label{fig:fig1}
\end{figure}

\textit{4DQH model and circuit realization.---}
The 4D lattice model is shown schematically in Fig.~\ref{fig:fig1}(a).  The spatial coordinates are denoted $x$, $y$, $z$, and $w$.  The lattice contains four sublattices labelled A, B, C and D, with sites connected by real nearest neighbour hoppings $\pm J$.  The four bands host two pairs of Dirac points in the Brillouin zone; each pair is the time-reversed counterpart of the other.  To control the pairs separately, long-range hoppings with amplitudes $\pm J'$ and $\pm J''$ are added within the $x$-$z$ plane [these long-range hoppings are omitted from Fig.~\ref{fig:fig1}(a) for clarity, but are shown in Fig.~\ref{fig:fig1}(c)].  Upon adding mass $+m$ to the A and B sites, and $-m$ to the C and D sites, the Dirac masses for the different Dirac point pairs close at $m\!=\!J'-2J''$ and $m\!=\!J''-2J'$. These gap closings are topological transitions, such that, for $J''\!=\!-J'$, the second Chern number of the lower bands is -2 (nontrivial) if $|m|\! <\! 3|J'|$.  Since T is unbroken, the first Chern number is always zero, so the model exhibits QH behaviour stemming purely from the second Chern number \cite{Price2018}.

For the experiment, we set $J=1$ and $J'=-J''=2$, so that the topological transition of the bulk lattice occurs at $m = \pm 6$.  We take a finite 4D lattice with three unit cells (6 sites) in the $x$ and $z$ directions, and one unit cell (2 sites) in $y$ and $w$.  Periodic boundary conditions are applied along $y$ and $w$ to mitigate finite-size effects, and are implemented using nearest neighbor type connections between opposite ends of the lattice.  The lattice has a total of 144 sites, of which we consider 16 to be bulk sites (defined as being more than 2 sites away from a surface) and 128 to be surface sites.

\textit{Circuit realization.---}
The finite 4D lattice is implemented with a set of connected printed circuit boards, shown in Fig.~\ref{fig:fig1}(b).  Each site $i$ of the tight-binding model maps to a node on the circuit, and the mass term maps to a circuit component of conductance $-D_{ii}$ connecting the node to ground.  Each hopping $J_{ij}$ between sites $i$ and $j$ maps to a circuit element of conductance $D_{ij}$ connecting the nodes. We add extra grounding components with conductance $D'_{ii}$ in parallel with $-D_{ii}$. If an external AC current $I_i$ flows into each node $i$ at frequency $f$, and $V_i$ is the complex AC voltage on that node, Kirchhoff's law states that
\begin{equation}
  I_i = (-D_{ii}+D'_{ii})V_i+\sum_{j}D_{ij}(V_i-V_j).
  \label{eq:Kirchhoff}
\end{equation}
We define
\begin{equation}
  \label{eq:mapping}
  D_{ij}(f)=i\alpha H_{ij}(f),
\end{equation}
where $\alpha$ is a positive real constant.  Then capacitances (inductances) correspond to positive (negative) real $H_{ij}$.  We require that at a reference working frequency $f = f_0$, the values of $H_{ij}(f_0)$ match the desired tight-binding lattice Hamiltonian.  We map the positive nearest neighbor hopping $J=1$ to capacitance $C_0=1\,\mathrm{nF}$ by taking $\alpha=2\pi f_0C_0$.  The long-range hopping $J'$ then maps to capacitance $C' = 2\,\mathrm{nF}$.  By setting $f_0 = 1/(2\pi\sqrt{L_0C_0}\,) \approx 113\,\mathrm{kHz}$, the negative nearest neighbor hopping maps to inductance $L_0 = 2\,\mathrm{mH}$, and the negative long-range hopping $J''=-2$ maps to inductance $L'=1\,\textrm{mH}$.

The grounding conductance of node $i$ is parameterised as $-D_{ii}+D'_{ii}$.  We tune $D_{ii}'$ so that for $f = f_0$ and $D_{ii}$ obeying Eq.~\eqref{eq:mapping}, $D'_{ii}+\sum_{j\neq i}D_{ij}=i\alpha E$ for a target energy $E$.  The required $D_{ii}'$ is dependent on the $m$ parameter.  Eq.~\eqref{eq:Kirchhoff} now becomes \cite{SM}
\begin{equation}
  I_i(f) \equiv \sum_{j}L_{ij}V_{j}
  = -i\alpha\sum_{j}\Big[H_{ij}(f) - E \, \delta_{ij}\Big] V_j(f),
  \label{eq:Kirchhoff1}
\end{equation}
where $L_{ij}$ are the components of the circuit Laplacian $L$.

In our experiments, we measure the impedance between a given node $r$ and the common ground by applying a $1\,\textrm{V}$ sine wave of frequency $f_0$ on that node, and measuring the voltage $V_r$ and the current $I_r$. The impedance between node $r$ and the ground is the $r$th diagonal term of the inverse of the circuit Laplacian $L$: 
\begin{align}
  \label{eq:ground_imp}
  V_r &= \sum_j(L^{-1})_{rj}I_j = Z_r I_r.
\end{align}
Using Eq.~\eqref{eq:Kirchhoff1}, one can show that \cite{Lee_2018}
\begin{align}	
  Z_r = \frac{i}{\alpha}\lim_{\epsilon \to 0}\sum_n
  \frac{|\psi_n(r)|^2}{E_n - E + i\epsilon},
  \label{eq:Zr}
\end{align}
where $\psi_n(r)$ is the $n$-th energy eigenstate's amplitude on site $r$, and $E_n$ is the corresponding eigenenergy.  Therefore $\mathrm{Re}[Z_r] = (1/\pi\alpha)\sum_n\delta(E-E_n) \, |\psi_n(r)|^2$ is, up to a scale factor, the LDOS of the target lattice at energy $E$ when measured at $f = f_0$.

\begin{figure}
\includegraphics[width=1.\columnwidth]{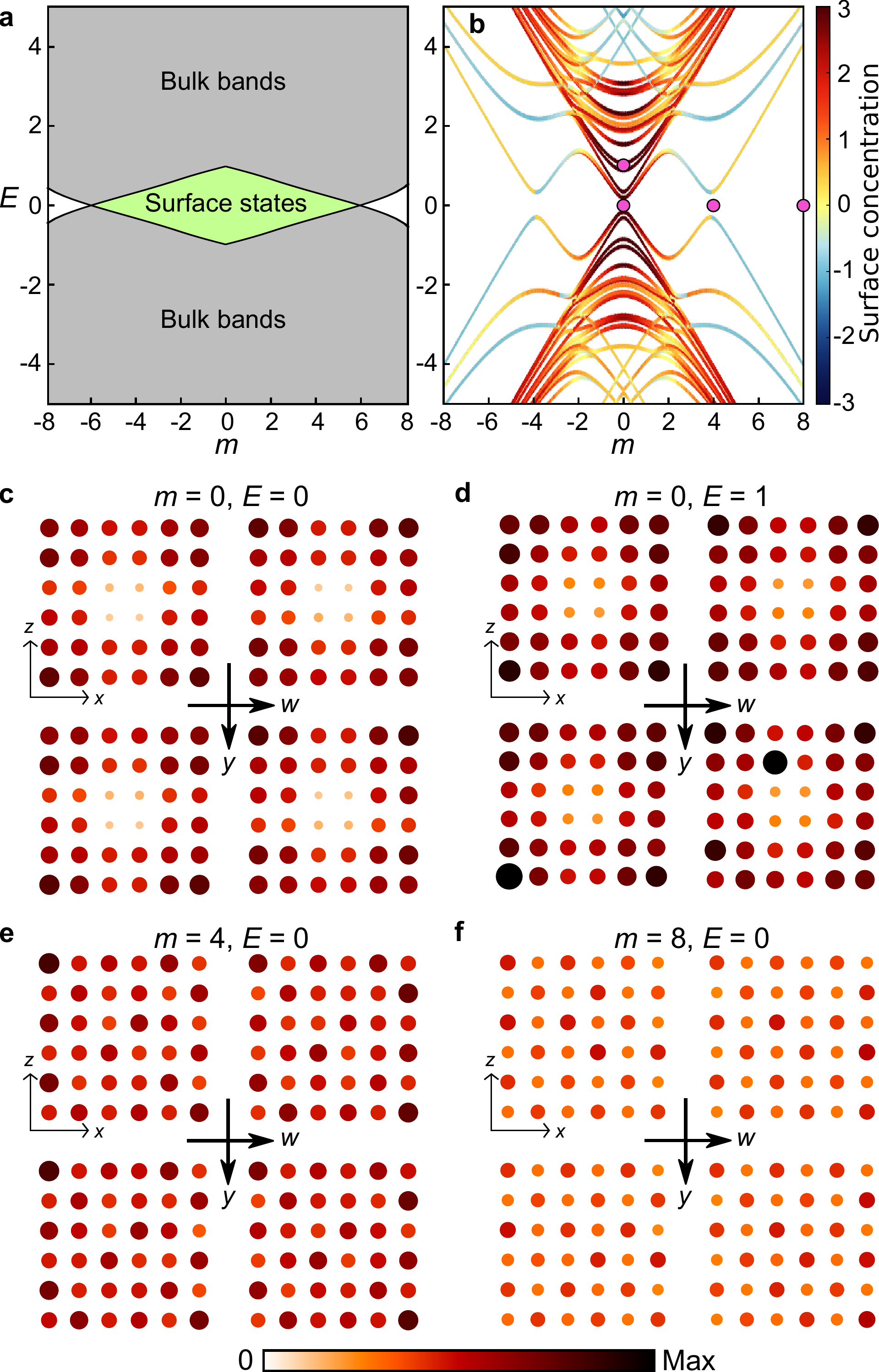}
\caption{(a) Calculated band diagram for the infinite 4D lattice.  The bulk bands are shown in gray.  For $|m| < 6$, there is a bandgap associated with nontrivial second Chern number, accompanied by topological surface states (shaded green).  For $|m| > 6$, the bandgap is trivial.  (b) Calculated band diagram for 144-site lattice with periodic boundary conditions along $y$ and $w$.  Colors indicate the degree of surface concentration of the energy states, as defined in the main text.  Due to finite-size effects, surface states occur at $|m| \lesssim 2$ and the gap closing is shifted to $|m| \approx 4$.  The parameters corresponding to subplots (c)--(f) are indicated with pink dots.  (c)--(f) Experimentally obtained LDOS maps for different $m$ and $E$, measured at working frequency $f = f_0$.  Surface states are observed in (c) and (d), consistent with theoretical predictions.}
\label{fig:fig2}
\end{figure}

\textit{Experimental results.---}
Fig.~\ref{fig:fig2}(a) shows the band diagram of the infinite bulk tight-binding model as a function of the mass detuning parameter $m$.  For $|m| < 6$, the system is in a 4DQH phase, with a topologically nontrivial bandgap centered at $E = 0$, which hosts topological surface states.  The band diagram for the 144-site tight-binding model is shown in Fig.~\ref{fig:fig2}(b).  The colors of the curves indicate the degree to which each eigenstate is concentrated on the surface, as defined by
\begin{equation}
  \ln\left[ \, \left\langle |\psi(r)| \right\rangle_{\mathrm{surf.}} \, / \,
    \left\langle |\psi(r)| \right\rangle_{\mathrm{bulk}} \,\right],
\end{equation}
where $\psi(r)$ denotes the energy eigenfunction, whose magnitudes are averaged over either surface or bulk sites.  Due to the finite lattice size, both the bulk and surface spectrum are split into sub-bands.  The closing of the bulk gap is shifted to $|m| \approx 4$, and the surface states occur most prominently at small values of $E$ and $|m|$.  In the Supplementary Information, we plot band diagrams for increasing lattice sizes, showing that the spectra come into close agreement with Fig.~\ref{fig:fig2}(a) as finite-size effects become negligible \cite{SM}.

\begin{figure*}
\includegraphics[width=2.0\columnwidth]{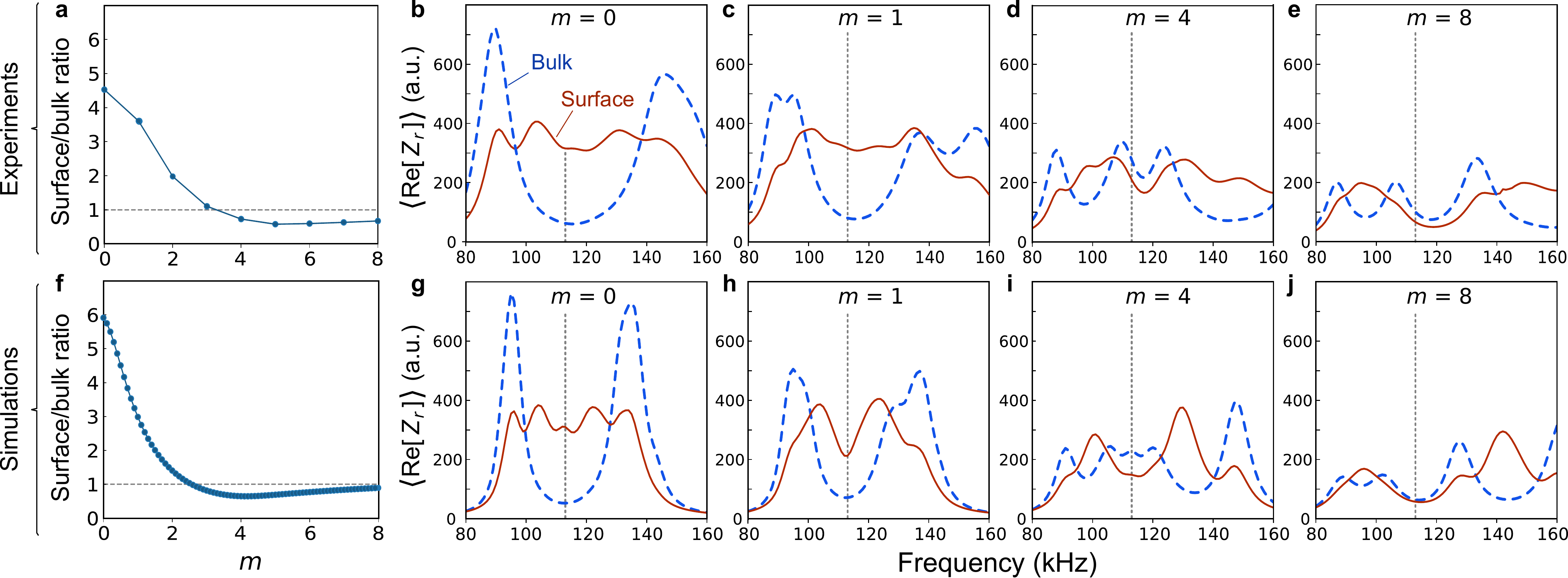}
\caption{Comparison of bulk and surface contributions to the LDOS.  (a) Ratio of surface to bulk LDOS, measured at $f = f_0$, versus $m$.  (b)--(e) Mean values of the LDOS measure $\mathrm{Re}[Z_r]$ on surface and bulk sites, versus working frequency $f$.  For these subplots, measurements were only taken over sites in the 2D plane $(y,w)=(1,0)$.  The reference working frequency $f_0$ (corresponding to $E = 0$) is indicated by the vertical dotted line.  For small $m$, we observe an elevated surface LDOS measure over a range of frequencies coincident with a bulk gap.  Upon increasing $m$, the gap closes.  (f)--(j) Simulation results corresponding to (a)--(e), computed using the same lattice size and circuit parameters.}
\label{fig:fig3}
\end{figure*}

We now fabricate a set of circuits with parameters $m \in \{0, 1, \dots, 8\}$ and $E \in \{0, 1\}$.  Figure~\ref{fig:fig2}(c)--(f) shows the measured LDOS (at $f = f_0$) for four representative samples.  From the experimental data, which agrees well with circuit simulations \cite{SM}, we see that the surface LDOS is high and the bulk LDOS is low when in the topologically nontrivial bandgap [Fig.~\ref{fig:fig2}(c) and (d)].  For $E = 0$, $m = 4$, which corresponds roughly to the gap-closing point, there is no significant difference between the surface and bulk LDOS.  For $E = 0$, $m = 8$, the LDOS on all sites is low, consistent with being in a topologically trivial bandgap.

To further quantify the differences between the 4DQH and conventional insulator phases, Fig.~\ref{fig:fig3}(a) plots the experimentally-determined ratio of the mean LDOS on surface sites to the mean LDOS on bulk sites, for different values of the mass detuning parameter $m$.  These measurements are performed at $f = f_0$, corresponding to $E = 0$.  With increasing $m$, the ratio decreases sharply from around $4.5$ in the 4DQH regime to around unity in the conventional insulator regime.  This result is consistent with the outcomes of circuit simulations [Fig.~\ref{fig:fig3}(f)].

The frequency dependent circuit impedances are consistent with the spectral features of a system with a topological phase transition.  Figure~\ref{fig:fig3}(b)-(e) plots the experimentally-obtained frequency dependence of the LDOS measure $\mathrm{Re}[Z_r]$, averaged over surface or bulk sites.  As explained above, our impedance measurements probe the response at fixed energy $E$ of an effective Hamiltonian $H(f)$ that depends parametrically on the frequency $f$ [Eq.~\eqref{eq:Zr}].  At the reference frequency, $H(f_0)$ matches the target tight-binding model; at other frequencies, $H(f)$ deviates from the form of the target model (e.g., the positive and negative hoppings become unequal in magnitude, lifting the band degeneracy).  However, so long as $E$ lies in the bulk bandgap of $H(f)$, the second Chern number is unchanged \cite{SM}.  Our experimental results at small values of $m$ indeed show a strong surface response over a range of frequencies spanning a bulk gap [Fig.~\ref{fig:fig3}(b) and (c)].  Increasing $m$ closes the bulk gap, and thereafter the surface and bulk LDOS measures exhibit no notable frequency dependent features.  These results also agree well with simulations [Fig.~\ref{fig:fig3}(g)--(j)].

\textit{Discussion.---}
We have used electric circuits to implement a 4D lattice hosting a 4D Quantum Hall phase.  This is the first experimental demonstration of a 4D topological lattice, and of a Class AI topological insulator. This is also the first experimental exploration of a 4DQH model with nontrivial second Chern number but trivial first Chern numbers.  Using impedance measurements, we have demonstrated that the LDOS on the 3D surface is enhanced in the 4DQH phase, due to the presence of topological surface states; the enhanced surface response is shown to span the frequency range of the bulk bandgap.  The gap closing is also clearly observed, despite being shifted by finite-size effects.  In future work, it is desirable to find ways to probe the detailed features of the 3D surface states, which are predicted to be two robust isolated Weyl points of the same chirality, a situation that does not occur in lower-dimensional topological models \cite{Price2018}.  This successful implementation of 4D lattices of very substantial size (144 sites) shows that electronic circuits are an excellent platform for exploring exotic band topological effects, and a promising alternative to the ``synthetic dimensions'' approach \cite{Boada_2012, Celi_2014, Yuan_2016} to realizing higher-dimensional lattices.

While this work was being done, we became aware of a related theoretical proposal to use circuits for realizing similar Class AI topological insulators \cite{Yu_arxiv}.

\begin{acknowledgments}
We are grateful to C.~H.~Lee and T.~Ozawa for helpful discussions.  This work was supported by the Singapore MOE Academic Research Fund Tier 3 Grant MOE2016-T3-1-006, Tier 1 Grants RG187/18 and RG174/16(S), and Tier 2 Grant MOE2018-T2-1-022(S).  HMP is supported by the Royal Society via grants UF160112, RGF/EA/180121 and RGF/R1/180071.
\end{acknowledgments}

\bibliography{biblist}
%
%
\begin{widetext}
\newpage

\makeatletter 
\renewcommand{\theequation}{S\arabic{equation}}
\makeatother
\setcounter{equation}{0}

\makeatletter 
\renewcommand{\thefigure}{S\@arabic\c@figure}
\makeatother
\setcounter{figure}{0}

\begin{center}
{\Large Supplementary Information}
\end{center}

\section{Supplementary Note 1: Circuit design details}

\begin{figure}[b]
\includegraphics[width=.55\columnwidth]{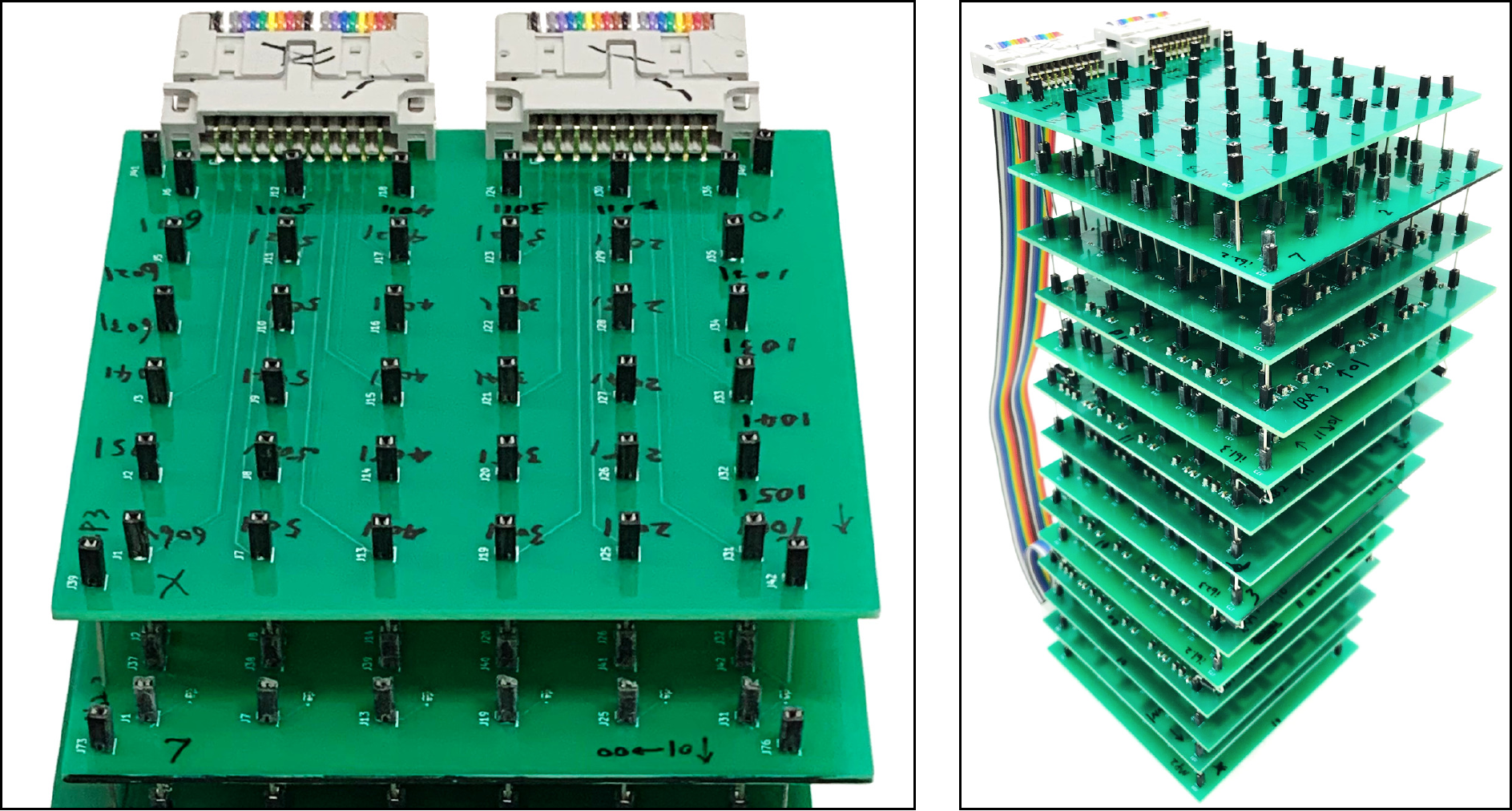}
\caption{Photographs showing the topmost PCB (left) and the stack of PCBs (right).}
\label{fig:figSphoto}
\end{figure}

The circuit is divided into several printed circuit boards (PCBs), stacked on top of each other.  As shown in Fig.~\ref{fig:figSphoto}, each PCB is divided into $6\times6 = 36$ nodes, corresponding to the dimensions of the 4D lattice in the $x$-$z$ plane; see Fig.~1(c) of the main text.  Each $x$-$z$ lattice plane actually consists of several PCBs stacked with vertical electrical interconnects, in order to fit all the necessary circuit components.

Let $I_i$ be the external current injected into node $i$, $V_j$ the voltage (relative to ground) on node $j$, and $D_{ij}$ the conductance between nodes $i$ and $j$ for $i\ne j$.  Moreover, let the conductance between node $i$ and ground be
\begin{equation}
  D_{ii}^{(g)} = -D_{ii} + D_{ii}'.
\end{equation}
By Kirchhoff's laws,
\begin{align}
  I_i &= D_{ii}^{(g)}V_i+\sum_{j}D_{ij}(V_i-V_j) \label{eq:SKirchhoff0} \\
  &= \sum_{j} \left[-D_{ij} + \left(-D_{ii} + D_{ii}'
    + \sum_k D_{ik}\right)\delta_{ij}\right] V_j \\
  &= \sum_{j} \left[-D_{ij} + \left(D_{ii}'
    + \sum_{k\ne i} D_{ik}\right)\delta_{ij}\right] V_j.
  \label{eq:SKirchhoff}
\end{align}
Note that in Eq.~\eqref{eq:SKirchhoff0}, the sum can be taken either over all $j$, or equivalently over $j\ne i$.  We now adjust $D_{ii}'$ so that, at a reference working frequency $f_0$,
\begin{equation}
  D_{ii}'(f_0) + \sum_{j\ne i} D_{ij}(f_0) = i \alpha E
  \label{SDtarget}
\end{equation}
for each node $i$, with some constant $\alpha$ and target energy $E$.  At $f = f_0$, Eq.~\eqref{eq:SKirchhoff} then becomes
\begin{align}
  I_i(f_0) &= -i\alpha \sum_{j} \Big[H_{ij}(f_0) - E\, \delta_{ij}\Big] V_j(f_0),
  \label{eq:SGreen} \\
  D_{ij}(f) &\equiv i\alpha H_{ij}(f).
\end{align}
We require $H_{ij}(f_0)$ to match the target tight-binding Hamiltonian, which has parameters $J = 1$, $J' = -J'' = 2$.  For real $\alpha$, positive (negative) real values of $H_{ij}$ correspond to capacitances (inductances).  As described in the main text, by choosing $\alpha$ and $f_0$ we can assign the following circuit elements to the lattice model's hopping terms:
\begin{align}
  \begin{aligned}
    C_0& &&\leftrightarrow\;\; J&\!\!\!\!&=1
    &&\textrm{(positive NN hopping)} \\
    C' &= 2C_0\;\; &&\leftrightarrow\;\; J'&\!\!\!\!&=2
    &&\textrm{(positive long range hopping)} \\
    L_0& &&\leftrightarrow -J&\!\!\!\!&=-1
    &&\textrm{(negative NN hopping)} \\
    L' &= L_0/2 &&\leftrightarrow\;\; J''&\!\!\!\!&=-2
    &&\textrm{(negative long range hopping)}
  \end{aligned}
  \label{eq:SCircuitElts}
\end{align}
where
\begin{equation}
  2\pi f_0 = 1/\sqrt{L_0 C_0}, \qquad \alpha = 2\pi f_0 C_0.
\end{equation}
    
For each node, we determine the grounding conductance required to satisfy Eq.~\eqref{SDtarget}.  Suppose node $i$ is connected to other nodes by $p_i$ type-$C_0$ capacitors, $q_i$ type-$L_0$ inductors, $p_i'$ type-$C'$ capacitors, and $q_i'$ type-$L'$ inductors (these connections depend on which sublattice the node lies on, and whether it lies in the bulk or on the surface).  Then
\begin{align}
  \begin{aligned}
  \sum_{j\ne i} D_{ij}(f) &= 2\pi i p_i f C_0 + \frac{q_i}{2\pi i f L_0}
  +  2\pi i p_i' f C' + \frac{q_i'}{2 \pi i f L'} \\
  &= 2 \pi i f\, C_0 \left(p_i+2p_i' - (q_i + 2q_i')
  \, \frac{f_0^2}{f^2} \right).
  \end{aligned}
  \label{eq:SsumDij}
\end{align}
Taking $f = f_0$ and plugging into Eq.~\eqref{SDtarget} gives
\begin{align}
  \begin{aligned}
  D_{ii}'(f_0) &=
  i \alpha E -\sum_{j\ne i} D_{ij}(f_0)\\
  &= 2\pi i f_0 C_0 \big(E - p_i - 2p_i' + q_i + 2q_i'\big).
  \end{aligned}
\end{align}
The on-site mass term is $H_{ii}(f_0) = \pm m$, depending on whether the node is on the A,B or C,D sublattices.  Hence, the grounding conductance must satisfy
\begin{align}
  \begin{aligned}
    D_{ii}^{(g)}(f_0) &= - D_{ii}(f_0) + D_{ii}'(f_0) \\
    &= 2 \pi i f_0 C_0 \big(E \mp m - p_i - 2p_i' + q_i + 2q_i'\big).
  \end{aligned}
  \label{equ:grounding_conductance}
\end{align}
To achieve this in the experiment, we connect each node $i$ to ground with $6-p_i$ type-$C_0$ capacitors, $3-q_i$ type-$L_0$ inductors, $4-p_i'$ type-$C'$ capacitors, and $4-q_i'$ type-$L'$ inductors.  Additionally, (i) we connect each node to ground by an extra inductor $L_g$, and (ii) if node $i$ belongs to sublattice C or D, we connect it to ground by an extra capacitor $C_m = 2mC_0$.  As a result, the grounding conductance of node $i$ at an arbitrary frequency $f$ is
\begin{align}
  D_{ii}^{(g)}(f) &=
  2\pi i (6-p_i) f C_0 + \frac{(3-q_i)}{2\pi i f L_0}
  + 2\pi i (4-p_i') f C' + \frac{(4-q_i')}{2 \pi i f L'}
  + \frac{1}{2 \pi i f L_g} + 2 \pi i (m \mp m) f C_0
\end{align}
where $\mp$ refers to sublattice A,B or C,D respectively.  At $f=f_0$, this satisfies Eq.~\eqref{equ:grounding_conductance} if we pick
\begin{equation}
  \frac{L_0}{L_g} = 3 + m - E.
\end{equation}
Hence,
\begin{equation}
  D_{ii}^{(g)}(f)  =
  2 \pi i f C_0\left[14 + m \mp m -p_i -2p_i'
    +\Big(E-14-m+q_i+2q_i'\Big)\frac{f_0^2}{f^2}
    \right].
  \label{eq:Diig}
\end{equation}

Returning to Eq.~\eqref{eq:SKirchhoff}, define the quantity in the parentheses---which gives rise to the $E$ term in Eq.~\eqref{eq:SGreen}---as
\begin{align}
  i\alpha\, \mathcal{E}_i(f) &= D_{ii}'(f) + \sum_{k\ne i} D_{ik}(f) \\
  &= D_{ii}^{(g)}(f) + D_{ii}(f) + \sum_{j\ne i} D_{ij}(f)
  \;\;\;\overset{f\rightarrow f_0}{\longrightarrow} \;\;i\alpha E.
  \label{eq:SiEi}
\end{align}
Eq.~\eqref{eq:SGreen} then generalises to
\begin{equation}
  I_i(f) = -i\alpha \sum_{j} \Big[H_{ij}(f)
    - \mathcal{E}_i(f)\, \delta_{ij}\Big] V_j(f).
  \label{eq:SGreen2}
\end{equation}
Now observe that in Eq.~\eqref{eq:SiEi}, the first term $D_{ii}^{(g)}(f)$ is defined by Eq.~\eqref{eq:Diig} for any $f$, and the third term is likewise defined by Eq.~\eqref{eq:SsumDij} for any $f$.  However, $D_{ii}(f)$ is defined only at $f = f_0$.  This turns out not to be a problem for our system of equations, since this term is exactly cancelled by the Hamiltonian term in Eq.~\eqref{eq:SGreen2}, which possesses the same ambiguity.  We are therefore free to give $D_{ii}(f)$ any frequency dependence, consistent with its value at $f_0$ (i.e., $D_{ii}(f_0) = i \alpha H_{ii}(f_0) = \pm i \alpha m$).  A convenient choice is
\begin{align}
  D_{ii}(f) &= i \alpha E - D_{ii}^{(g)}(f) - \sum_{j\ne i} D_{ij}(f)
  \;\;\;\overset{f\rightarrow f_0}{\longrightarrow} \;\;i\alpha H_{ii}(f_0)
  \label{eq:Sondiag} \\
  \Rightarrow\quad
  i\alpha\, \mathcal{E}_i(f) &= i \alpha E \quad \textrm{for all}\;\;i, f.
\end{align}
With this choice, $\mathcal{E}_i(f)$ becomes $i$-independent, and Eq.~\eqref{eq:SGreen2} simplifies to
\begin{equation}
  I_i(f) = -i\alpha \sum_{j} \Big[H_{ij}(f) - E\, \delta_{ij}\Big] V_j(f).
  \label{eq:SGreen3}
\end{equation}
This can be interpreted as a family of response equations with an $f$-dependent Hamiltonian and fixed energy $E$.  For general $f$, the Hamiltonian's hopping terms are determined by the circuit elements summarised in Eq.~\eqref{eq:SCircuitElts}, and its on-site mass term is determined by Eq.~\eqref{eq:Sondiag}; for $f = f_0$, it reduces to the target Hamiltonian.

Suppose $E$ is in a topological gap of the target Hamiltonian, so that topological edge states exist at frequency $f_0$.  As we vary $f$ away from $f_0$, the Hamiltonian varies smoothly, deviating from the form of the target Hamiltonian (e.g., the positive and negative nearest neighbor hoppings become unequal in magnitude).  Throughout this variation, so long as $E$ lies in a gap, the topological properties are unchanged and the topological edge states continue to exist.  Thus, the $f$-dependent response of the circuit behaves like a bandstructure.  For small $m$, the circuit exhibits a finite-width topological bandgap in $f$-space; tuning $m$ closes this bandgap, and causes the topological edge states to disappear.

\section{Supplementary Note 2: finite-size effects in the tight-binding model}

\begin{figure}[b]
\includegraphics[width=.5\columnwidth]{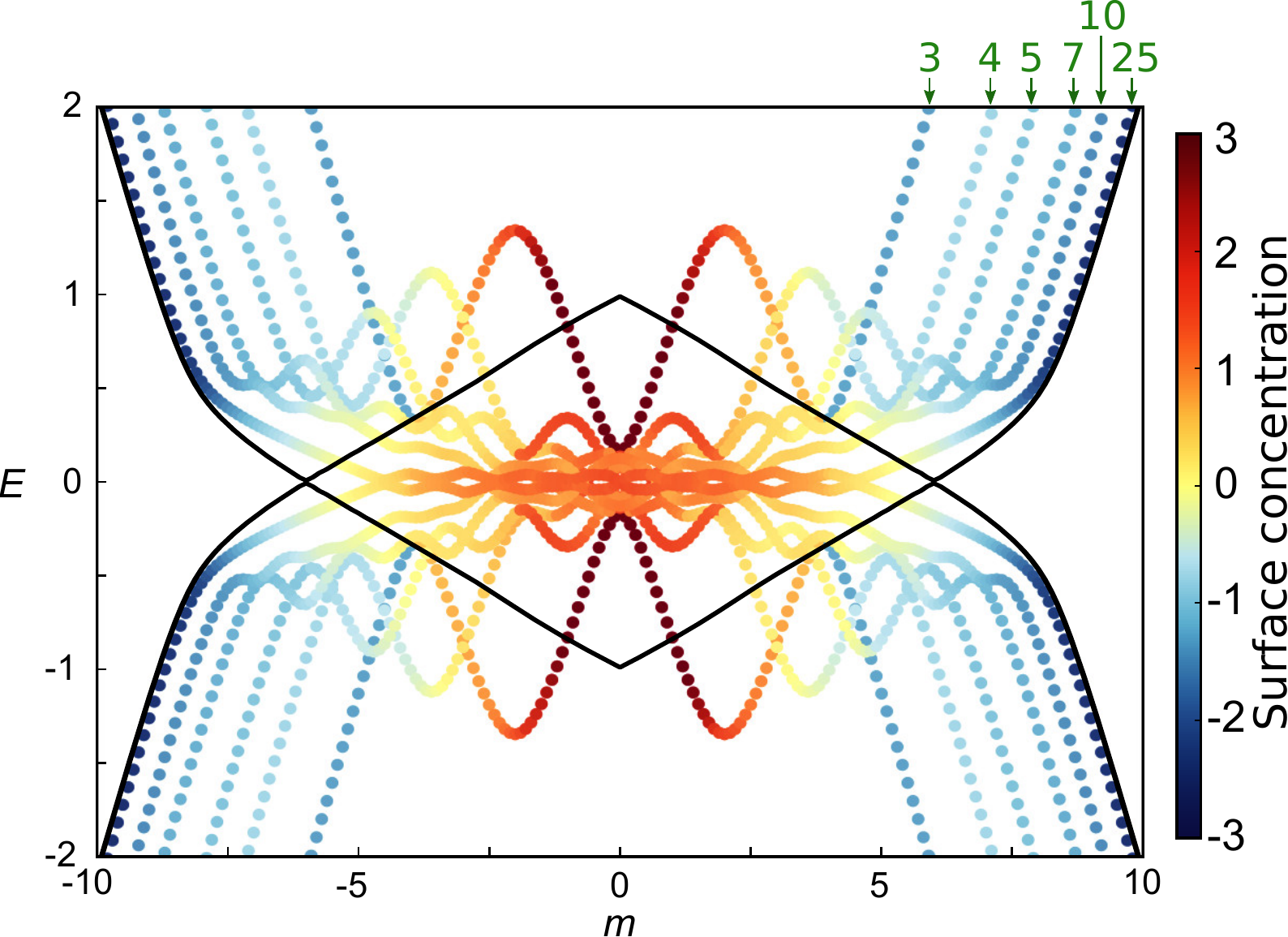}
\caption{Eigenvalue pairs closest to $E = 0$ for lattices with $\{3, 4, 5, 7, 10, 25\}$ unit cells along $x$, $z$ (lengths labelled in the upper right corner), and one unit cell along $y$, $w$ (with periodic boundary conditions).  Black curves show the bulk band edges.} 
\label{fig:figSPD}
\end{figure}

As shown in Fig.~1(a) of the main text, the bandgap of the infinite tight-binding model closes at $m=6$.  However, as shown in Fig.~1(b), for a finite lattice of the same size as our experimental sample the gap closing occurs around $m=4$; moreover, the edge states appear at small values of $m$.

To show that this discrepancy is a standard finite-size effect, Fig.~\ref{fig:figSPD} plots the band edges (i.e., the pair of eigenvalues closest to $E = 0$) for a series of lattices with $\{3, 4, 5, 7, 10, 25\}$ unit cells along both $x$ and $z$.  Along $y$ and $w$, these lattices remain one unit cell wide with periodic boundary conditions (equivalent to $k_y = k_w = 0$).  The colors indicate whether the eigenstate is concentrated on the surface (red) or in the bulk (blue).  As the size of the lattice increases, the eigenvalues at large $m$ (in the conventional insulator regime) approach the predicted bulk band edges, while the eigenvalues in the topological insulator regime spread over a larger range of $m$ corresponding to the topologically nontrivial gap.

\section{Suplementary Note 3: Frequency response for different mass parameters}

\begin{figure}[b]
\includegraphics[width=.55\columnwidth]{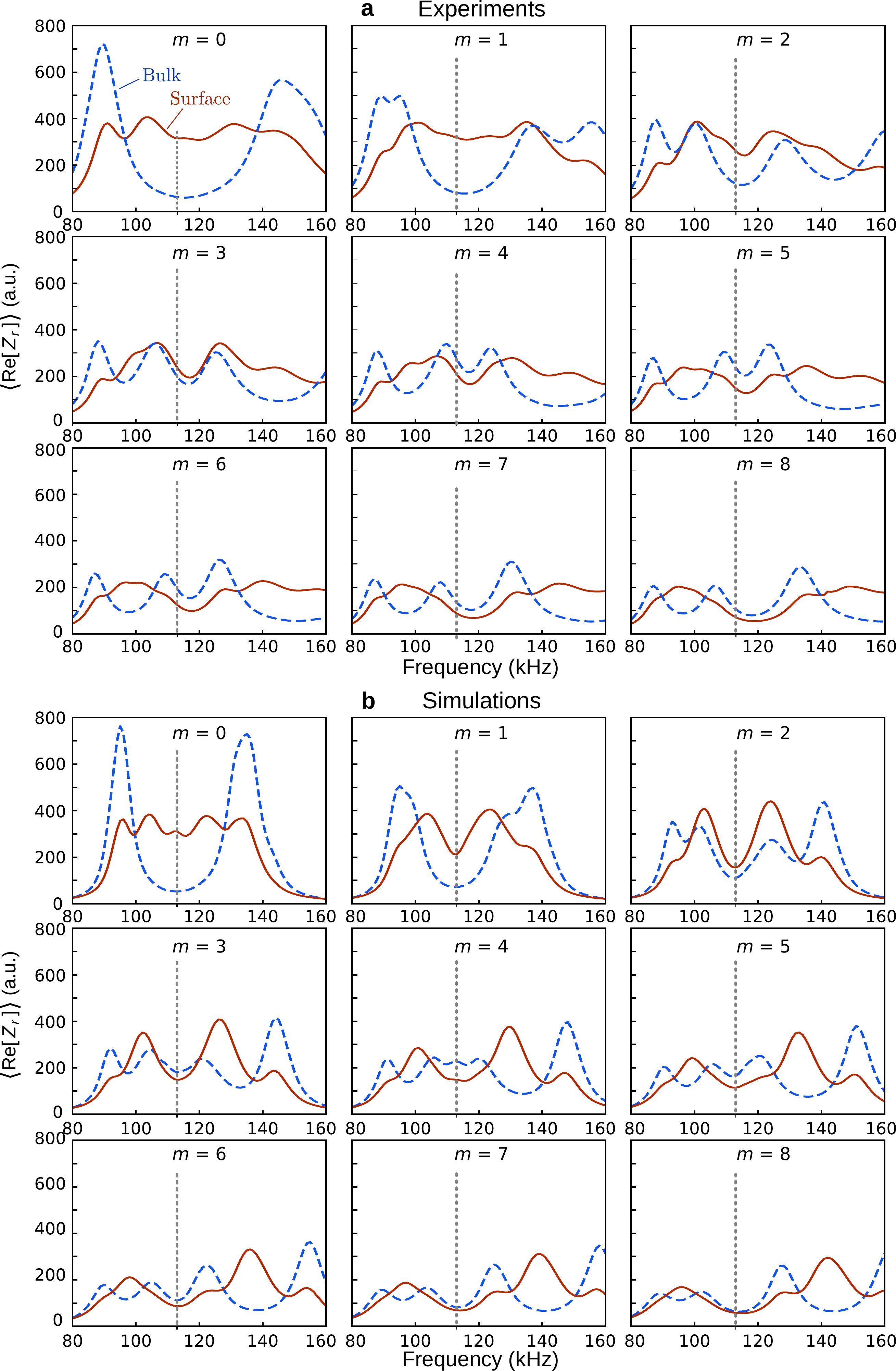}
\caption{(a) Measured frequency dependence of $\mathrm{Re}[Z_r]$ averaged over surface and bulk sites, for $m = 0, 1, \dots, 8$.  The measurements are taken over sites in the 2D plane $(y,w)=(1,0)$, and $f_0$ is indicated by the vertical dotted lines.  (b) The corresponding circuit simulation results.}
\label{fig:figSfreq}
\end{figure}

Fig.~\ref{fig:figSfreq} plots the frequency dependence of the LDOS measure $\mathrm{Re}[Z_r]$, averaged over on surface and bulk sites, for $m = 0, 1, \dots, 8$.  In Fig.~3(b)--(e) and (g)--(j), only a few values of $m$ were shown for brevity.  The experimental results are shown in Fig.~\ref{fig:figSfreq}(a), and the corresponding simulation results are shown in Fig.~\ref{fig:figSfreq}(b).  The discrepancies between experimental and simulation results can be explained by the disorder in the experimental samples: according to the manufacturer data sheets, individual capacitors and inductors have 10\% tolerance in the stated capacitances and inductances; moreover, as discussed in Supplementary Note 4, there are variations in the resistances of the individual capacitors, inductors, and interconnects.

\section{Supplementary Note 4: Details of circuit simulations}

All circuit simulations are performed with \texttt{ngspice}, a free software circuit simulator.  To model circuit losses, we treat each capacitor and inductor as having a $5 \Omega$ resistance in series with the purely capacitive or inductive element.  This has the same order of magnitude as the resistances stated in the data sheets for the individual circuit components; we pick a uniform value of $5 \Omega$ to represent the various hard-to-characterise resistances in the PCBs.

The simulations are performed like the experiments: i.e., we apply a sine wave voltage source to each node, use the steady-state voltage and current to determine the complex impedance, and hence obtain the LDOS on each site.  In Fig.~\ref{fig:SLDOSsim}, we show how the simulated frequency-dependent LDOS measure is affected by circuit resistance.  The upper row shows the outcomes for $5\Omega$ resistances, identical to the results shown in Fig.~\ref{fig:figSfreq} and Fig.~3 of the main text (which are a good match for experimental results).  The lower row shows the results with an order of magnitude lower resistance ($0.5\Omega$).  As can be seen, the main effect of the circuit resistances is to smooth out the frequency dependence of the LDOS measure, but the signatures of the topological edge states and bulk bandgap are present in either case.

Figure~\ref{fig:figSLDOSsim} shows the simulated LDOS profile for $E=0, m=0,4,8$ and $E=1, m=0$ (using $5\Omega$ resistances), corresponding to the experimental results shown in Fig.~2(c)--(f) of the main text.

\begin{figure}[h]
\includegraphics[width=\columnwidth]{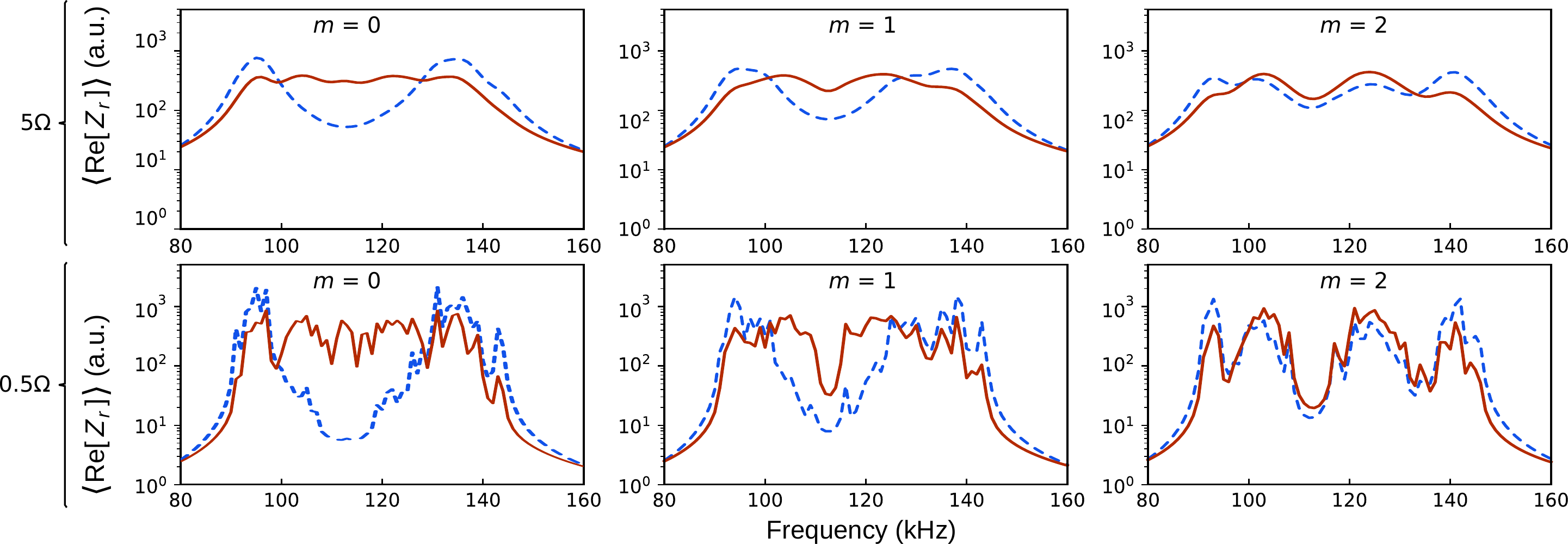}
\caption{Simulated frequency-dependent LDOS measure for different circuit component resistances: $5\Omega$ (upper row), corresponding to the first row of Fig.~\ref{fig:figSfreq}(b), and $0.5\Omega$ (lower row).}
\label{fig:SLDOSsim}
\end{figure}

\begin{figure}[h]
\includegraphics[width=\columnwidth]{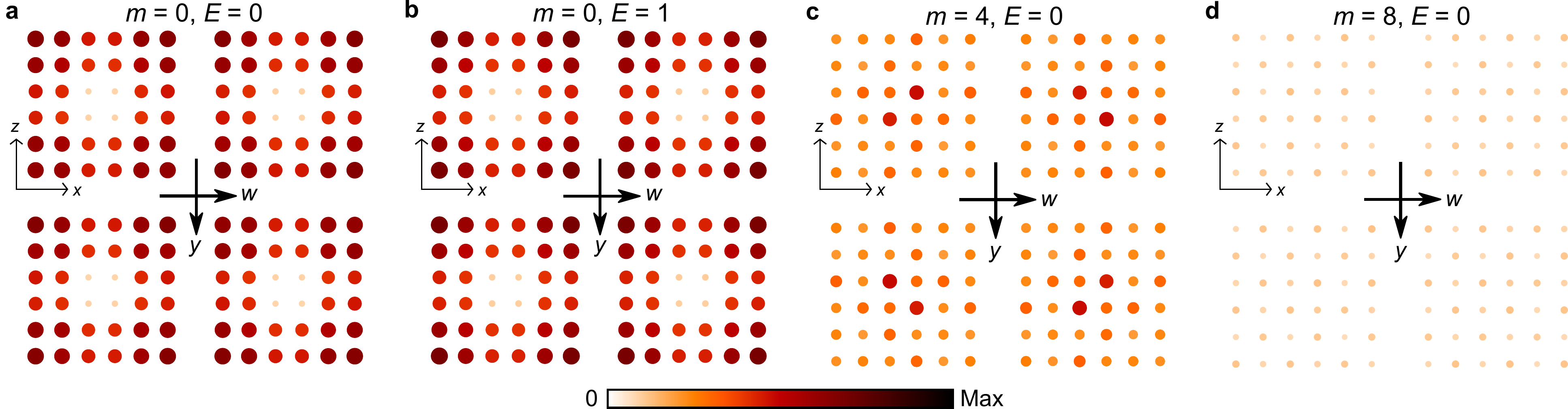}
\caption{Simulated LDOS distribution for $E=1,m=0$ and $E=0,m=0,4,8$; compare to Fig.~2(c)--(f) of the main text.}
\label{fig:figSLDOSsim}
\end{figure}

\clearpage
\end{widetext}
\end{document}